\documentstyle [12pt] {article}
\title{THE SIMPLEST DETERMINATION OF THE THERMODYNAMICAL
     CHARACTERISTICS OF KERR-NEWMAN BLACK HOLE}

\author{Vladan Pankovi\'c$^{\ast,\sharp}$, Simo Ciganovi\'c$^\sharp$,
Rade Glavatovi\'c$^\diamond$\\
$^\ast$Department of Physics, Faculty of Sciences, 21000 Novi
Sad,\\ Trg Dositeja Obradovi\'ca 4. , Serbia, vdpan@neobee.net \\
$^\sharp$Gimnazija, 22320 Indjija, Trg Slobode 2a, Serbia\\
$^\diamond$ Military-Medical Academy, 11000 Belgrade, Crnotravska
17., Serbia \\}

\date {}
\begin{document}
\maketitle


\begin {abstract}
In this work, generalizing our previous results, we determine in
an original and the simplest way three most important
thermodynamical characteristics (Bekenstein-Hawking entropy,
Bekenstein quantization of the entropy or (outer) horizon surface
area and Hawking temperature) of Kerr-Newman black hole. We start
physically by assumption that circumference of Kerr-Newman black
hole (outer) horizon holds the natural (integer) number of
corresponding reduced Compton's wave length and use
mathematically, practically, only simple algebraic equations. (It
is conceptually similar to Bohr's quantization postulate in Bohr's
atomic model interpreted by de Broglie relation.)
\end {abstract}
\vspace {0.5cm}
"Here have been deceased clouds\\
dead time with history of days,\\
here have been fallen rays;\\
nirvana oppressed whole universe."\\
 \vspace{0.2cm}
 \hspace{3cm}   Vladislav Petkovi\'c Dis ,"Nirvana"
\vspace {0.5cm}

In this work, generalizing our previous results [1], [2], we shall
reproduce and determine in the  simplest way three well-known
[3]-[9], most important thermodynamical characteristics
(Bekenstein-Hawking entropy, Bekenstein quantization of the
entropy or (outer) horizon surface area and  Hawking temperature)
of  Kerr-Newman black hole. We shall start  physically by
assumption that circumference of black hole (outer) horizon holds
the natural (integer) number of corresponding reduced Compton's
wave length and use mathematically, practically, only simple
algebraic equations. (It is conceptually similar to Bohr's
quantization postulate in Bohr's atomic model interpreted by de
Broglie relation.)  In this work we shall use natural system of
the units where speed of light, Planck constant, Newtonian
gravitational constant and Boltzmann constant are equivalent to
unit.

Consider a Kerr-Newman black hole with mass $M$, angular momentum
$J$, electrical charge $Q$, outer horizon radius
\begin {equation}
    R_{+}= M + (M^{2} - {\it a}^{2} - Q^{2})^{\frac {1}{2}}
\end {equation}
and surface area of the outer horizon
\begin {equation}
    A_{+} = 4\pi(R^{2}_{+} + {\it a}^{2})
\end {equation}
for
\begin {equation}
    {\it a} = \frac {J}{M}
\end {equation}

Suppose the following expression
\begin {equation}
      m_{+n}R_{+} = \frac {n}{2\pi}, \hspace{1cm}  {\rm for} \hspace{0.3cm} n = 1,
      2,....
\end {equation}
that implies
\begin {equation}
      2\pi R_{+} = n \frac {1}{m_{+n}}=n\cdot\lambda_{r+n} \hspace{1cm}{\rm for} \hspace{0.3cm} n = 1,
      2,...
\end {equation}
Here $2\pi R_{+}$ represents the circumference of the outer
horizon while
\begin {equation}
      \lambda_{r+n}= \frac {1}{m_{+n}}
\end {equation}
represents the $n-th$ reduced Compton wavelength of a quantum
system with mass $ m_{+n}$ captured at black hole outer horizon
for $n = 1, 2,...$ .

Expression (5) simply means that {\it circumference of the black
hole outer horizon holds exactly}$n$ {\it corresponding} $n$-{\it
th reduced Compton wave lengths of  quantum systems for} $n = 1,
2,...$ . Obviously, it is conceptually similar to well-known
Bohr's quantization postulate interpreted by de Broglie relation
(according to which circumference of $n$-th electron circular
orbit contains exactly $n$ corresponding $n$-th de Broglie wave
lengths, for $n = 1, 2,...$).

According to (4) it follows
\begin {equation}
       m_{+n} = n\frac {1}{2\pi R_{+}}= m_{+1} , \hspace{1cm}{\rm for} \hspace{0.3cm}  n = 1,
       2,...
\end {equation}
where
\begin {equation}
       m_{+1} = \frac {1}{2\pi R_{+}}
\end {equation}
Obviously, $ m_{+1}$ depends of $M$ so that $ m_{+1}$ decreases
when $M$ increases and vice versa. For a macroscopic black hole,
i.e. for $M \gg 1$ it follows $ m_{+1}\ll 1$.

Now, we shall define the following variable
\begin {equation}
       M_{+} = S_{+} m_{+1} = \frac { A_{+}}{4} m_{+1} =  \frac {1}{2}\frac{ R^{2}_{+}+{\it a}^{2} }{ R_{+}}
\end {equation}
where $ S_{+}$ represents the Bekenstein-Hawking entropy
\begin {equation}
     S_{+} = \frac { A_{+}}{4} =  \pi  (R^{2}_{+}+{\it a}^{2})            .
\end {equation}
As it is not hard to see it follows
\begin {equation}
   M_{+}\simeq M  \hspace{0.9cm}{\rm for} \hspace{0.3cm} M^{2}\gg  {\it a}^{2} + Q^{2}
\end {equation}
and
\begin {equation}
   M_{+} = \frac {M}{2} (1 + \frac {{\it a}^{2}}{ M^{2}}) \ge \frac {M}{2} \hspace{1cm} {\rm for} \hspace{0.3cm}M^{2} =  {\it a}^{2} + Q^{2}                 .
\end {equation}

According to (9) it follows
\begin {equation}
   S_{+} = \frac { M_{+}}{ m_{+1}}
\end {equation}
which can give a rough intuitive presentation of the black hole
entropy. Namely, since according to usual (classical) definition
\begin {equation}
   S_{+} = \ln N
\end {equation}
where $N$ represents the number of the statistical microstates, it
follows
\begin {equation}
  N = \exp \frac { M_{+}}{ m_{+1}}
\end {equation}
which represents a rough (classical) estimation of the number of
the statistical microstates within given black hole.

Now, we shall differentiate (10) under some additional
supposition. Namely, we shall formally, i.e. approximately suppose
that by given differentiation all terms that hold a can be
considered almost constant so that their derivations can be
neglected. It yields
\begin {equation}
    d S_{+}= 2\pi  R_{+}d R_{+}
\end {equation}
Further, it will be supposed that formally, i.e. approximately it
is satisfied
\begin {equation}
    d R_{+} = (1 + \frac {M}{(M^{2} - {\it a}^{2} - Q^{2})^{\frac {1}{2}}}) dM = \frac { R_{+}}{(M^{2} - {\it a}^{2} - Q^{2})^{\frac {1}{2}}}dM  \simeq 2dM        .
\end {equation}
Obviously, both (16) and (17) can be satisfied for $ M^{2}\gg
{\it a}^{2} + Q^{2}$.

Introduction of (17), in (16), yields
\begin {equation}
dS_{+} = 2 \pi \frac { R^{2}_{+}}{(M^{2} - {\it a}^{2} -
Q^{2})^{\frac {1}{2}}}dM  \simeq 4\pi
  R_{+}dM
\end {equation}
or, in a corresponding finite difference form
\begin {equation}
    \Delta S_{+} = 4\pi R_{+} \Delta M \hspace{1cm} {\rm for} \hspace{0.3cm}   \Delta M \ll M  .
\end {equation}
Further, we shall assume
\begin {equation}
   \Delta M = n m_{+1} \hspace{1cm} {\rm for} \hspace{0.3cm} n = 1, 2, …
\end {equation}
which, according to (1), (8), yields
\begin {equation}
    \Delta S_{+} = 2n \hspace{1cm} {\rm for} \hspace{0.3cm}  n = 1, 2, …
\end {equation}
that represents the Bekenstein quantization of the black hole
entropy. It, according to (10), implies
\begin {equation}
    \Delta A_{+} = 8n = 2n (2)^{2} \hspace{1cm} {\rm for} \hspace{0.3cm}  n = 1, 2, …
\end {equation}
that represents the Bekenstein quantization of the black hole
surface, where $2^{2}$ represents the surface of the quadrate
whose side length represents twice Planck length, i.e. 1.

Now we shall attempt to determine T+ in the following way. We
shall consider first thermodynamical law for Kerr-Newman black
hole
\begin {equation}
   dM = T_{+}dS_{+} +  \Omega_{+}dJ +  \Phi_{+}dQ
\end {equation}
where
\begin {equation}
    \Omega_{+} = \frac {{\it a}}{ R^{2}_{+} + {\it a}^{2}}
\end {equation}
represents the outer horizon rotation rate, i.e. angular speed,
and,
\begin {equation}
    \Phi_{+}= Q \frac { R_{+}}{ R^{2}_{+} + {\it a}^{2}}
\end {equation}
- outer horizon electrostatic potential.

Further, we shall approximately neglect term $\Phi_{+} dQ$ in (23)
so that this expression turns out in
\begin {equation}
  dM = T_{+}dS_{+} +  \Omega_{+}dJ
\end {equation}
According to (3) it follows
\begin {equation}
    J = {\it a} M
\end {equation}
so that, according to previous supposition, i.e. approximate
condition that a represents a constant, it follows
\begin {equation}
     dJ = {\it a} dM        .
\end {equation}

Introduction of accurate form of (17), and  (24), (27), (29) in
(26) yields the following algebraic equation with unknown variable
$T_{+}$
\begin {equation}
     1 = T_{+}2 \pi \frac { R^{2}_{+}}{(M^{2} - {\it a}^{2} - Q^{2})^{\frac {1}{2}}}dM  + \frac {{\it a}^{2}}{R^{2}_{+} + {\it a}^{2}} .
\end {equation}
It yields
\begin {equation}
     T_{+} = \frac {1}{2\pi} \frac {(M^{2}- {\it a}^{2} - Q^{2})^{\frac {1}{2}}}{R^{2}_{+} + {\it a}^{2}}
\end {equation}
that represents the Hawking temperature for Kerr-Newman black
hole.

In this way we have reproduced, i.e. determined exactly, in the
simplest way, three most important thermodynamical characteristics
of Kerr-Newman black hole: Bekenstein-Hawking entropy (13),
Bekenstein quantization of the black hole entropy (21) or
Bekenstein quantization of the black hole surface area (22), and,
Hawking temperature (31).

It can be shortly repeated and pointed out that our results are
done starting,  physically, by assumption that circumference of
Kerr-Newman black hole (outer) horizon holds the natural (integer)
number of corresponding reduced Compton's wave length, and,
mathematically, practically, by simple algebraic equations only.
All this is conceptually similar to Bohr's quantization postulate
in Bohr's atomic model interpreted by de Broglie relation. Roughly
speaking, we gave simply effectively exact predictions on the
three well-known most important thermodynamical characteristics of
Kerr-Newman black hole (Bekenstein-Hawking entropy, Bekenstein
quantization of the black hole entropy or Bekenstein quantization
of the black hole surface area, and, Hawking temperature) from a
"mesoscopic", like to quasi-classical, view point. But of course,
many other important characteristics of Kerr-Newman black hole
cannot be obtained by our simple model. However, our predictions
are in the excellent agreement with Copeland and Lahiri work [10].
Namely, Copeland and Lahiri, starting from "microscopic", i.e.
string theory, demonstrated that thermodynamical characteristics
of the (Schwarzschild) black hole can be obtained by a standing
waves corresponding to small oscillations on a circular loop with
radius equivalent to (Schwarzschild) horizon radius.

\vspace {0.5cm} Authors are very grateful to Prof. Dr. Petar
Gruji\'c for illuminating discussions. He represents, really and
unambiguously, the very coauthor of this work. Also, authors are
very grateful to Prof. Dr. Darko Kapor and Prof. Dr. Miodrag Krmar
for many (infinite) different forms of the support and help.

\vspace {0.5cm} This work is dedi{\it cat}ed to memory of Mi\'ca
Mali.

\vspace{1cm}

{\large \bf References}

\begin {itemize}

\item [[1]] V. Pankovic, M. Predojevic, P.Grujic,
{\it A Bohr's Semiclassical Model of the Black Hole Thermodynamics}, gr-qc/0709.1812
\item [[2]] V. Pankovic, J. Ivanovic, M. Predojevic, A.-M. Radakovic,
{\it The Simplest Determination of the Thermodynamical
Characteristics of Schwarzschild, Kerr and Reisner-Nordstr$\ddot
{o}$m black hole}, gr-qc/0803.0620
\item [[3]] J. D. Bekenstein, Phys. Rev., {\bf D7}, (1973), 2333.
\item [[4]] S. W. Hawking, Comm. Math. Phys., {\bf 43}, (1975), 199
\item [[5]] S. W. Hawking, Phys. Rev., {\bf D14}, (1976), 2460.
\item [[6]] S. W. Hawking, in {\it General Relativity, an Einstein Centenary Survay}, eds. S. W. Hawking, W. Israel (Cambridge University Press, Cambridge, UK 1979)
\item [[7]] R. M. Wald, {\it Black Hole and Thermodynamics}, gr-qc/9702022
\item [[8]] R. M. Wald, {\it The Thermodynamics of Black Holes}, gr-qc/9912119
\item [[9]] D. N. Page, {\it Hawking Radiation and Black Hole Thermodynamics}, hep-th/0409024
\item [[10]] E. J. Copeland, A.Lahiri, Class. Quant. Grav. , {\bf 12} (1995) L113 ; gr-qc/9508031

\end {itemize}

\end {document}